\documentclass[twocolumn]{aastex631}
\usepackage{graphicx}
\usepackage{graphbox}
\usepackage{hyperref}
\usepackage{amssymb,amsmath}
\usepackage{stmaryrd}
\usepackage{booktabs}
\usepackage{orcidlink}

\usepackage[print-unity-mantissa=false, separate-uncertainty=true, multi-part-units=single]{siunitx}
\usepackage{tikz}
\usetikzlibrary{shapes.geometric, arrows.meta, positioning}

\DeclareSIUnit\erg{erg}
\DeclareSIUnit\arcsecond{as}
\DeclareSIUnit\year{yr}
\DeclareSIUnit\Jansky{Jy}

\newcommand{\lea}{\mathrel{<\kern-1.0em\lower0.9ex\hbox{$\sim$}}}
\newcommand{\lta}{{\>\rlap{\raise2pt\hbox{$<$}}\lower3pt\hbox{$\sim$}\>}}
\newcommand{\gta}{{\>\rlap{\raise2pt\hbox{$>$}}\lower3pt\hbox{$\sim$}\>}}

\begin{document}
\title{The Intermediate Mass Black Hole in Omega Centauri: Constraints on Accretion from JWST}

\shorttitle{IMBH in Omega Centauri}

\author{Steven Chen\orcidlink{0000-0001-6169-2731}}
\affiliation{Department of Physics, The George Washington University, 725 21st St. NW, Washington, DC 20052}
\email{schen70@gwmail.gwu.edu}

\author{Jeremy Hare\orcidlink{0000-0002-8548-482X}}
\affiliation{NASA Goddard Space Flight Center, Greenbelt, MD, 20771}
\affiliation{NASA Postdoctoral Program Fellow}

\author{Oleg Kargaltsev\orcidlink{0000-0002-6447-4251}}
\affiliation{Department of Physics, The George Washington University, 725 21st St. NW, Washington, DC 20052}

\author{Hui Yang\orcidlink{0000-0002-8832-6077}}
\affiliation{IRAP-Roche, 9, avenue du Colonel Roche BP 44346 31028 Toulouse France}

\author{Denis Cioffi}
\affiliation{Department of Physics, The George Washington University, 725 21st St. NW, Washington, DC 20052}

\author{Maximilian H\"{a}berle\orcidlink{0000-0002-5844-4443}}
\affiliation{European Southern Observatory, Karl-Schwarzschild Strasse 2 85748, Garching, Germany}

\author{Anil Seth\orcidlink{0000-0003-0248-5470}}
\affiliation{Department of Physics, The University of Utah, 201 Presidents Circle, Salt Lake City, UT 84112}

\begin{abstract}

\noindent We analyze JWST observations of the central region of the globular cluster $\omega$ Centauri (NGC 5139, $\omega$ Cen hereafter), around the position of the candidate IMBH inferred by \cite{haberle_fast-moving_2024} from the motion of fast-moving stars in multi-epoch HST observations. We performed PSF-fitting photometry for sources in NIRCam (F200W and F444W) and MIRI (F770W and F1500W) and constructed UV to IR SEDs for sources within the central region of the cluster by using HST photometry from oMEGACat \citep{haberle_omegacat_2024}. None of the SEDs of reliably measured sources within this region resembles the SEDs computed from models of \cite{pesce_toward_2021} for IMBHs accreting from intracluster medium at low rates. Our JWST limits place constraints on combinations of IMBH mass and accretion rate, either due to the amount of material available to be accreted, or due to the fraction of accreting matter that actually falls into the IMBH. Our non-detection then does not necessarily contradict the mass range of the IMBH inferred from the fast moving stars. We discuss these constraints in the context of the model of \cite{pesce_toward_2021}. We find that JWST limits are more restrictive than the existing radio limits for IMBH masses $\lesssim 6000 M_{\odot}$. It is also possible that the faint IMBH emission is dominated by the light of a nearby star. Tighter limits on accretion onto the candidate IMBH can be placed with deeper observations, a more precise localization of the IMBH, and better measurements of the local intracluster medium density and temperature at the center of the cluster. 

\end{abstract}


\section{Introduction}

Observed black holes (BHs) are commonly divided into stellar mass BHs, with masses $\lesssim \qty{100}{M_\odot}$ and supermassive BHs, with masses$\gtrsim \qty{1e5}{M_\odot}$ \citep{kormendy_coevolution_2013}. Intermediate-mass black holes (IMBHs), with masses ranging $\sim\qtyrange{1e3}{1e5}{M_\odot}$, have much more scant observational evidence, although they would provide the missing link between stellar mass BHs and supermassive BHs \citep{kremer_compact_2025}. The centers of globular clusters (GCs) are long thought to host IMBHs, but only recently has strong observational evidence for such objects been found in GCs, both galactic and extragalactic, e.g. \cite{mezcua_observational_2017, romero-shaw_gw190521_2020, greene_intermediate-mass_2020}. 

Searches for IMBHs in GCs can involve detection of emission from the IMBH, or observations of the impact of the IMBH on GC dynamics. Models of cluster evolution suggests that IMBHs can lead to cusps in the surface density and alter the velocity dispersion profiles of clusters\citep{umbreit_monte_2012,umbreit_constraining_2013,baumgardt_n_2017,greene_intermediate-mass_2020,evans_dark_2022}, or even alter the mass segregation profile at the core \citep{gill_intermediate-mass_2008,umbreit_constraining_2013}. These methods were used, for example, by \cite{umbreit_monte_2012} and \cite{umbreit_constraining_2013} to constrain the mass of an IMBH in NGC 5694 to $\lesssim 1000 M_\odot$ and in M10 to $\lesssim 600 M_\odot$. Dynamical evidence of IMBHs have sometimes been controversial, for example in 47 Tuc, where initially an IMBH with $M\sim2,200M_{\odot}$ was reported by \cite{kiziltan_intermediate-mass_2017}, but was then disputed by \cite{mann_multimass_2019} and \cite{henault-brunet_black_2020}. More recently, \cite{paduano_ultradeep_2024} found a radio source at the center of 47 Tuc that may be a millisecond pulsar, or an IMBH of up to $6000 M_\odot$. Constraints for 35 other clusters based on N-body simulations are reviewed in \cite{baumgardt_n_2017} with a typical limit on the IMBH mass being a few thousand $M_{\odot}$. Constraints based on upper limits of radio and X-ray emissions of various GCs are also presented in \cite{tremou_maveric_2018} and \cite{su_chandra_2022}, respectively.

Omega Centauri (NGC\,5139, $\omega$~Cen hereafter) has long been a prime target for IMBH searches. Early star velocity dispersion and surface-brightness cusp measurements suggested a central black hole of $M_{\rm BH} \sim (3$–$5)\times10^{4}\,M_\odot$ \citep{noyola_gemini_2008}. Subsequent HST proper-motion work that revised the cluster center and modeled the stellar kinematics in the cluster's core placed a tighter bound at $1\sigma$ limits of $M_{\rm BH} \lesssim 1.2\times10^{4}\,M_\odot$ \citep{anderson_new_2010,van_der_marel_new_2010}. Dynamical degeneracies with radial anisotropy and/or a centrally concentrated population of stellar-mass black holes were shown to further reduce the need for an IMBH to explain the velocity dispersion in the core \citep{zocchi_radial_2017}. More recently, a combined analysis of stellar kinematics and line-of-sight accelerations of millisecond pulsars instead produced evidence for an extended central dark mass ($\sim 2$–$3\times10^{5}\,M_\odot$) and set a $3\sigma$ upper limit $M_{\rm BH}<6\times10^{3}\,M_\odot$ on the IMBH mass \citep{banares-hernandez_new_2025}. 

Stronger recent evidence for an IMBH in $\omega$~Cen has been presented by \cite{haberle_fast-moving_2024} (H+24 hereafter) based on a new, high-precision proper-motion catalog (oMEGACat; \citealt{haberle_omegacat_2024}). The unprecedented depth and precision of this catalog enabled the discovery of seven faint, fast-moving stars within the central 3\arcsec of the cluster (0.08 pc at the cluster distance of 5.49 kpc \citep{haberle_omegacat_2025}). The proper motions (PMs) of these stars correspond to velocities that exceed the cluster's escape velocity of $\approx$62~km~s$^{-1}$. According to H+24, the presence of these stars can only be explained if they are gravitationally bound to a central, massive, and compact object. From the kinematics of the five most robustly measured fast-moving stars alone, H+24 inferred a firm lower limit on the black hole's mass of approximately 8,200~M$_{\odot}$. Further analysis incorporating acceleration limits and detailed N-body simulations suggests a mass range of $39,000-47,000$~M$_{\odot}$. The search for these stars was centered on the cluster's position determined by \cite{anderson_new_2010} (hereafter, the ``AvdM10'' center), located at R.A. = $13^{\mathrm{h}}26^{\mathrm{m}}47.24^{\mathrm{s}}$ (201.69683$^{\circ}$), Dec. = $-47^{\circ}28'46.45''$ ($-47.47957^{\circ}$) (J2000). This detection provides one of the strongest cases to date for an IMBH in the local universe. \cite{prieto_growing_2025} provided theoretical backing through Monte-Carlo N-body models of the evolution of $\omega$ Cen. These models are able to reproduce the growth of an IMBH to a mass of $50,000$~M$_{\odot}$, similar to the range reported by H+24, as well as a population of fast-moving stars. 

Direct searches of IMBH emissions in $\omega$ Cen have been attempted at radio and X-ray wavelengths. \cite{haggard_deep_2013} did not detect IMBH emission with the \textit{Chandra X-ray Observatory}, and obtained X-ray luminosity limits of $<\qty{1.7e30}{\erg\per\second}$. With their accretion model, they concluded that the putative IMBH has a mass less than a few thousand $M_\odot$. The MAVERIC survey \citep{tremou_maveric_2018} searched for IMBH radio emission in $\omega$ Cen using the Australia Telescope Compact Array (ATCA), and obtained radio flux limits of $<\qty{8.9}{\micro\Jansky}$ at 5 GHz. Assuming IMBHs accrete from intracluster gas in a manner similar to low-luminosity AGN, they reported an even more stringent mass limit of $<\qty{1000}{M_\odot}$ at the $3\sigma$ significance level. Both works concluded that if the IMBH is more massive than $\qty{10000}{M_\odot}$, then it must be an extraordinarily inefficient accretor. 
Since the submission of this work, deeper ATCA radio limits on the IMBH were published by \cite{mahida_no_2025}, achieving $3\sigma$ upper limits of $\qty{3.3}{\micro\Jansky}$ at 5.5 GHz. We use this updated upper limit in our analysis. However, we still refer to \cite{tremou_maveric_2018}, since \citep{mahida_no_2025} do not provide their own estimates of the IMBH mass, and the analysis frameworks of both papers are similar. 

Recent work by \cite{pesce_toward_2021} (hereafter P+21) and \cite{seepaul_detectability_2022} (hereafter SPN22) suggests that direct detection of IMBH emission is more feasible in the infrared (IR) for low accretion rates. The latter paper focuses specifically on detectability of IMBHs in our Galaxy, while the former describes the details of the Spectral Energy Distribution (SED) calculation. In particular, Figure 2 of SPN22 shows that at low accretion rates ($\dot{M}/\dot{M}_{\mathrm{Edd}} \sim 10^{-8}$), the IR luminosity of IMBHs is higher than the radio, optical, and X-ray luminosity, (see also \citealt{murchikova_observability_2025}). This motivated us to carry out a sensitive search for a faint EM counterpart of an IMBH in the core of $\omega$ Cen with the James Webb Space Telescope (JWST). Cross-matching existing HST data to the JWST data allows for SEDs to be constructed from ultraviolet (UV) through IR frequencies, which can be used to distinguish any IMBH candidates from normal stars. 

In Section \ref{processing}, we describe the HST and JWST data used, and outline the data reduction, alignment, and cross-matching procedures. We present images and color magnitude diagrams (CMDs) of the sources in the central region of $\omega$ Cen in Section \ref{results}, and analyze sources with an apparent red excess. This is followed by calculations of detection limits within the central region in Section \ref{upper limits}. Based on our results, we discuss IMBH mass limits in $\omega$ Cen inferred using the P+21 radiative model in Section \ref{Discussion} and conclude with a summary of our findings in Section \ref{conclusions}. 

\section{Observations and Data Processing}
\label{processing}

\subsection{HST Catalog}
\label{HST data}

A new, extensive HST catalog of $\omega$ Cen, oMEGACat, was presented in \cite{haberle_omegacat_2024}. This catalog provides deep and high-precision data products covering the cluster's inner region out to its half-light radius of approximately 5\arcmin. This includes positions and proper motions relative to $\omega$ Cen for 1.4 million stars observed with the Advanced Camera for Surveys Wide Field Camera (ACS/WFC) and Wide Field Camera 3 (WFC3) UVIS channel onboard HST. The measurements are based on a temporal baseline covering a $\approx$20-year period between 2002 and 2023 (with most observations of the central region concentrated between 2010 and 2023), which allows for exceptionally high precision measurements. For stars with a magnitude of $m_{\text{F625W}} \approx 18$, the median one-dimensional proper-motion error is $\approx \qty{11}{\micro\arcsecond\per\year}$. In the most densely observed central region ($r < 1.5\arcmin$), the precision is even better, reaching $\sim \qty{6.6}{\micro\arcsecond\per\year}$. The reference frame used in oMEGACat is comoving with the cluster center, which has coordinates frozen at the epoch of 2002.5, such that background sources (e.g., galaxies) are moving in the sky instead. The positions of cluster stars, which move due to their individual motions in the cluster, are given at the epoch of 2012.0. 

oMEGACat provides precise photometry in seven filters: ACS/WFC F435W, F625W, F658N and WFC3/UVIS F275W, F336W, F814W. The corresponding images fully cover the $\sim5\arcmin$ half-light radius of $\omega$ Cen. An additional WFC3/UVIS filter, F606W, has extensive coverage in the central region, but not in the outer regions. The photometric data have been empirically corrected for spatially dependent variations caused by differential reddening and instrumental effects. The photometry is reliable for stars down to $m_{\text{F625W}} \approx 25$ mag. The catalogs, as well as deep, stacked images for each of the seven filters are available from oMEGACat's Zenodo repository \citep{haberle_data_2024}\footnote{\url{doi:10.5281/zenodo.11104046}}. 

\subsection{JWST Observations}
\label{JWST observations}

JWST observed the central region of $\omega$ Cen on July 10, 2024 (MJD 60501) with NIRCam\footnote{\url{https://jwst-docs.stsci.edu/jwst-near-infrared-camera\#gsc.tab=0}} and on July 29, 2024 (MJD 60520) with MIRI\footnote{\url{https://jwst-docs.stsci.edu/jwst-mid-infrared-instrument\#gsc.tab=0}} for program GO-4343 (PI: O. Kargaltsev). NIRCam imaging observations were taken with module B and used the F200W and F444W filters with scientific exposures of 515s in each filter. An INTRAMODULEBOX dither pattern was employed, to fill gaps in sky coverage between detectors, together with the BRIGHT1 readout pattern. MIRI imaging observations were taken with the F770W and F1500W filters with scientific exposures of 1,887 s and 6,394 s, respectively. A 4-point dither pattern and the FASTR1 readout pattern was utilized. All required data products were downloaded from Mikulski Archive for Space Telescopes (MAST)\footnote{At the time of download, the calibration pipeline version was 1.16.1, and the CRDS version was 12.0.5 \citep{bushouse_jwst_2024}}.

\subsection{JWST Data Reduction}
\label{dolphot}

We used the NIRCam and MIRI modules in the DOLPHOT pipeline to detect sources and perform point-spread function (PSF) fitting photometry on the JWST images \citep{weisz_jwst_2024}. The preprocessing steps in the pipeline include the masking of bad pixels and the creation of a sky image in each filter. After preprocessing, DOLPHOT takes as input a reference image (an \textsc{I2D} file that is a rectified and calibrated image mosaic\footnote{\url{https://jwst-docs.stsci.edu/accessing-jwst-data/jwst-science-data-overview\#gsc.tab=0}}), a list of science images (\textsc{cal} files, on which DOLPHOT performs the PSF-fitting photometry), and a set of parameters. The parameters include settings to determine image alignment, source detection, photometry modes, as well as significance thresholds for a source to be retained in the final catalog. DOLPHOT fits PSF models to signal-to-noise peaks simultaneously on all images of one instrument. 

To run the NIRCam module, we followed the steps described in the ``User’s guide for DOLPHOT NIRCam and NIRISS modules\footnote{\url{https://dolphot-jwst.readthedocs.io}}.'' The NIRCam DOLPHOT parameters were set to values specified in the M92 example\footnote{\url{https://github.com/ers-stars/dolphot_jwst/blob/main/docs/running_dolphot/m92_NIRCam_phot.param}} (\citealt{weisz_jwst_2024}). For the MIRI module, we used the parameter settings recommended in the ``DOLPHOT/MIRI User’s Guide''\footnote{\url{http://americano.dolphinsim.com/dolphot/dolphotMIRI.pdf}} for crowded fields, which differ from the default parameter values by using \textsc{FitSky} = 2, \textsc{img RAper} = 3, and \text{img RSky2} = 4 10. For the reference images, we used the drizzled F200W image (jw04343-o002\_t001\_NIRCam\_clear-f200w\_i2d) for NIRCam, and F770W image (jw04343-o001\_t00\_miri\_f770w\_i2d) for MIRI. 

The final outputs of DOLPHOT are separate NIRCam and MIRI source catalogs ($\sim 1$ million and $\sim 66$ thousand sources respectively), each of which contains photometric information from each filter. Since our goal is the detection of a possibly very faint and red source, we use a completeness-oriented set of relaxed photometric quality cuts to select NIRCam sources (SNR $\geq 10$, Crowding $\leq2.25$, Sharpness$^2\leq0.1$, Object Type $\leq2$, Photometry Quality Flag $\leq3$ \footnote{see \url{https://dolphot-jwst.readthedocs.io/en/latest/post-processing/catalogs.html}}). All cuts are applied to each NIRCam filter, as well as to the global NIRCam photometry, which combines information from each filter. As a result of these loose cuts, many spurious sources, or sources with contaminated flux measurements, in the streaks of bright stars and in especially crowded regions are not removed. However, this approach minimizes the chances of discarding any real sources. We do not apply any cuts on MIRI photometric quality, as the MIRI PSF is significantly worse (0.269\arcsec for MIRI F770W, vs. 0.062\arcsec for NIRCam F200W), and even such loose cuts appear to remove many MIRI counterparts to NIRCam sources. 

\subsection{Alignment to oMEGACat}
\label{astrometry}

Searching for a potential IMBH requires sub-arcsecond level cross-matching between the oMEGACat, NIRCam, and MIRI catalogs. We propagate the oMEGACat catalog to the epoch of the JWST observations, and then correct for residual astrometric shifts using the \textsc{TweakReg} module from \textsc{DrizzlePac}\footnote{\url{https://drizzlepac.readthedocs.io/en/latest/index.html}}. The details of the alignment procedure can be found in Appendix \ref{appendix_alignment}; here we mention major steps. \textsc{TweakReg} computes a transformation that aligns coordinates from an input catalog to coordinates in a reference catalog. While transformations can include shifts, rotations, and scaling, we only use shift transformations as they are sufficient for this work. First, we propagated oMEGACat coordinates to the JWST epoch using both the absolute GC PM ($\mu_\alpha \cos\delta = \qty{-3.2510}{\milli\arcsecond\per\year}$, $\mu_\delta = \qty{-6.7398}{\milli\arcsecond\per\year}$), and individual star PMs relative to the cluster PM, computed by H+24. We then aligned the NIRCam catalog to the PM-propagated oMEGACat catalog ($\Delta_{RA} = 0.014\arcsec$, $\Delta_{DEC} = 0.036\arcsec$) and the MIRI catalog to the corrected NIRCam catalog ($\Delta_{RA} = -0.193\arcsec$, $\Delta_{DEC} = 0.063\arcsec$). The absolute GC PM was also used to propagate the WCS in the oMEGACat images to the JWST epoch, and the computed transformations were used to update the WCS in the corresponding NIRCam and MIRI images to produce aligned images. We use these images and catalogs throughout the rest of the paper. 

\subsection{Cross-matching Sources Around Cluster Center}
\label{crossmatching}

We cross-match the oMEGACat, NIRCam, and MIRI catalogs in the regions within $r=1\arcsec$, 3\arcsec, and 40\arcsec of the ``AvdM10'' center of $\omega$ Cen used in H+24. After propagating to the JWST epoch, the ``AvdM10'' center has coordinates of RA$=201.696806^\circ$ and Dec$=-47.479614^\circ$. The $r<1\arcsec$ region contains both of the two estimates of the position of the IMBH, as well as four of the fast moving stars identified by H+24. However, since the estimate of the MCMC IMBH position has $1\sigma$ uncertainties of the order of 1\arcsec, we also search the $r<3\arcsec$ region around the center. We discuss sources with an apparent red excess (see Figure \ref{fig:ngc5139_center_cmds}) within the $r<1\arcsec$ and 3$\arcsec$ regions in Section \ref{CMDs}. The $r<40\arcsec$ region covers most of the MIRI field-of-view, and is used to compare sources found in the smaller regions against a larger reference sample.

\begin{figure*}[h]
    \centering
    \includegraphics[width=0.5\textwidth]{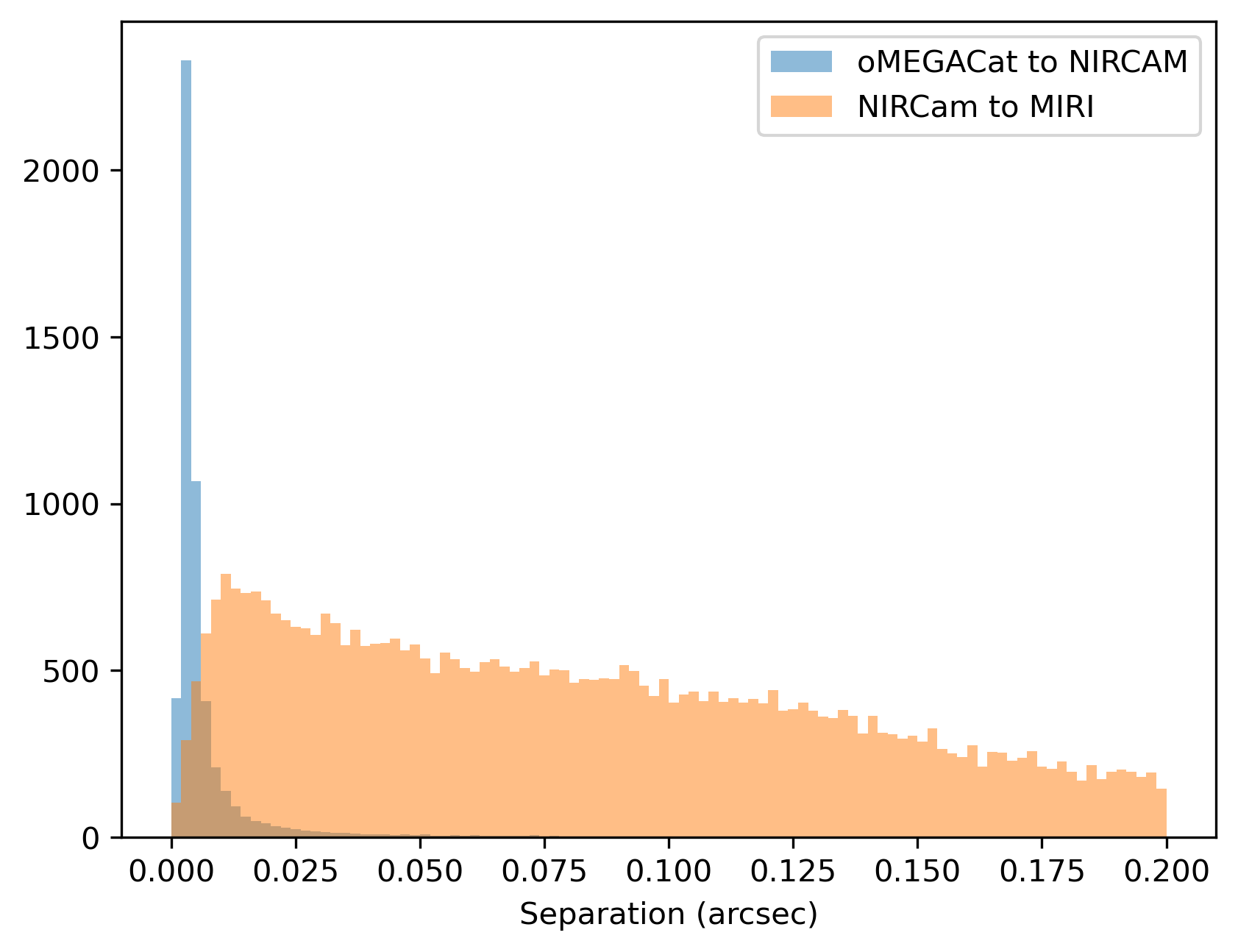}
    \caption{Distribution of the separation to the nearest match between PM-propagated oMEGACat and NIRCam, and NIRCam and MIRI sources.}
    \label{fig:ngc5139_d2d_jwst_oMEGACat}
\end{figure*}

Figure \ref{fig:ngc5139_d2d_jwst_oMEGACat} shows the distribution of the separation of the nearest oMEGACat to NIRCam, and MIRI to NIRCam matches for the 40\arcsec region without any photometry cuts. Most NIRCam matches are less than 0.01\arcsec away from the oMEGACat position, speaking to the high quality of the alignment solution. However, the MIRI distribution is significantly more spread out -- up to $\sim 0.1\arcsec$, indicating the presence of more numerous spurious sources, or faint sources that were blended and detected as a single MIRI source. 

Based on Figure \ref{fig:ngc5139_d2d_jwst_oMEGACat}, we first match each oMEGACat source to the closest NIRCam source within 0.02\arcsec, and then match the corresponding NIRCam source to the closest MIRI source within 0.1\arcsec. We then identify all NIRCam sources lacking oMEGACat counterparts and match them to the closest MIRI source within 0.1\arcsec. Lastly, we identify sources that only appear in MIRI. Since many MIRI sources are likely to be blended detections of multiple adjacent NIRCam sources, we allow a single MIRI source to be matched to multiple NIRCam sources within 0.1\arcsec. Around 18\% of MIRI sources in the 40\arcsec region are matched to multiple NIRCam sources within 0.1\arcsec.

\section{Results}
\label{results}

\subsection{Images}
\label{images}

\begin{figure*}[h]
    \centering
    \includegraphics[width=0.49\textwidth]{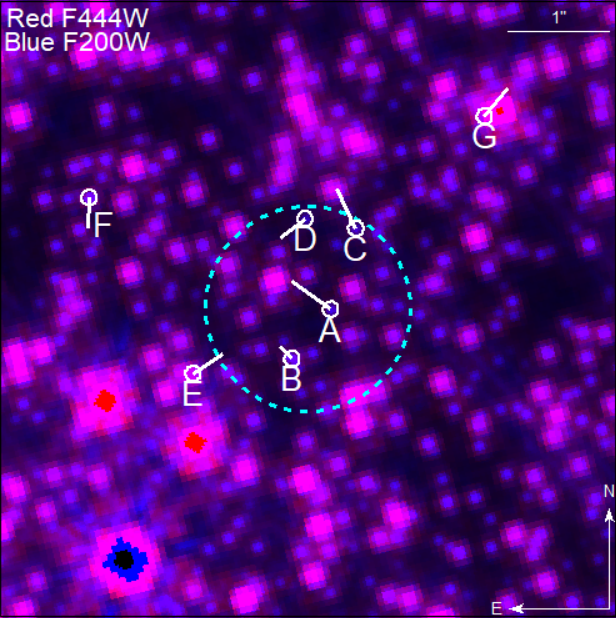}
    \includegraphics[width=0.49\textwidth]{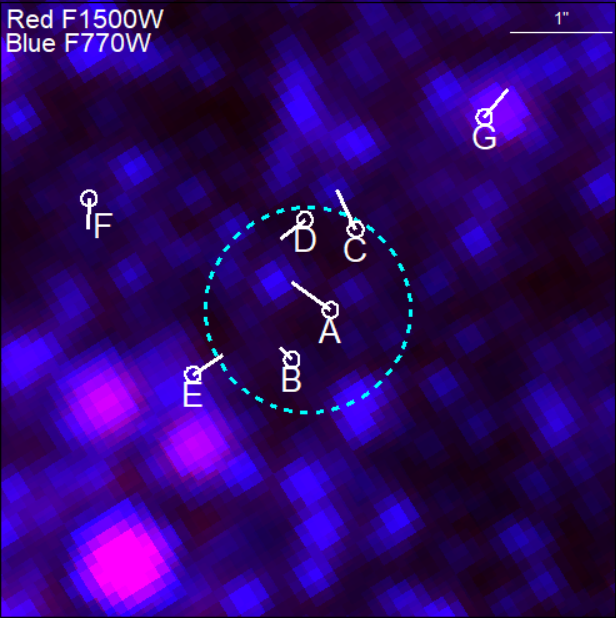}
    \includegraphics[width=0.49\textwidth]{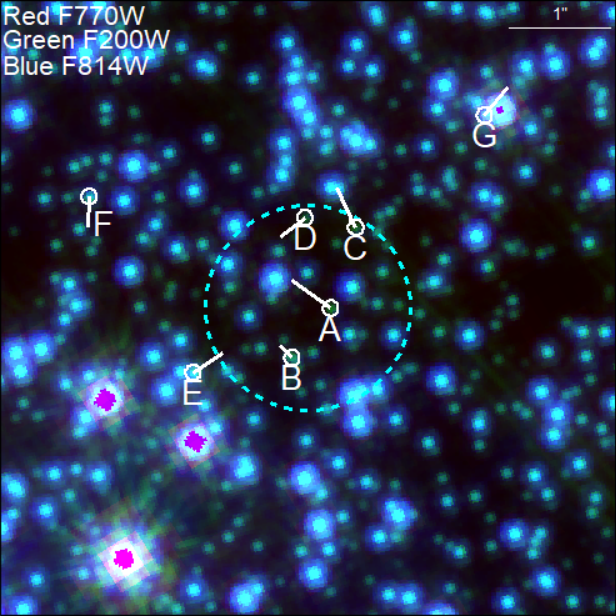}
    \includegraphics[width=0.49\textwidth]{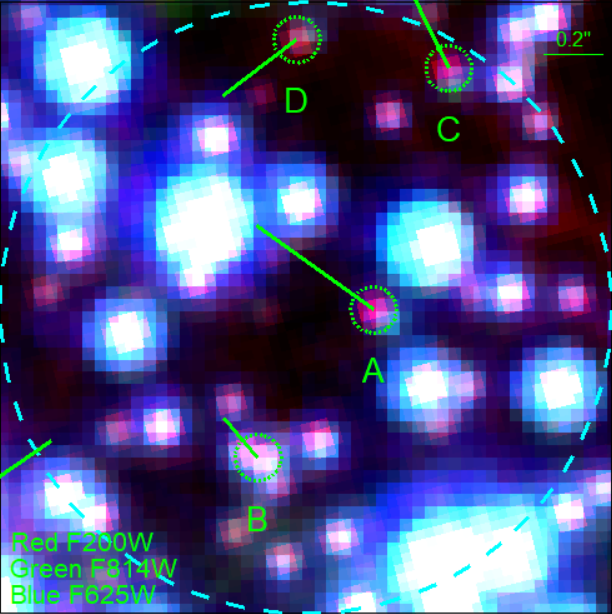}
    \caption{False color images of the IMBH RoI, centered on the same position (after PM propagation) as Figure 1 from H+24. Fast moving stars are labeled as in H+24, with their oMEGACat to NIRCam PMs, scaled to 120 years, shown with lines. The red and blue colors correspond to NIRCam F444W and F200W images in the top left panel, and to MIRI F1500W and F770W images in the top right panel. The RGB colors correspond to MIRI F770W, NIRCam F200W, and HST F814W images in the bottom left panel, and to NIRCAM F200W, HST F814W and HST F625W images in the bottom right panel. The first three panels show $6\arcsec \times 6\arcsec$ regions around the cluster center, while the last panel zooms in on the $2\arcsec \times 2\arcsec$ region to show the PMs of the fast moving stars. The dashed cyan circle with $r=1''$ is centered at the ``AvdM10" cluster center. 
    }
    \label{fig:ngc5139_center_fast_stars}
\end{figure*}

False color images of the cluster's center region ($6''\times6''$ and $2''\times2''$) are shown in Figure \ref{fig:ngc5139_center_fast_stars} for several combinations of HST and JWST filters. The fast moving stars in H+24 are labeled in the top left panel. Since the images were aligned by propagating the oMEGACat images to the JWST epoch using the absolute cluster PM (unlike the catalogs, which were aligned using both cluster and star PMs), individual PMs of the fast moving stars are discernible in the zoomed-in bottom right image as an offset in the source position between different filters. 
We computed the PMs of these stars by comparing their oMEGACat and NIRCam positions in the reference frame of the cluster. These PMs (shown in Figure \ref{fig:ngc5139_center_fast_stars} as lines extrapolated to 120 years from JWST observation epoch) qualitatively agree well with those from H+24, except stars B and G, which lie near other stars and thus suffer substantial contamination. There are several other stars with visibly high PMs that were not listed in H+24. We defer more accurate measurements of PMs and an analysis of their kinematics to a future publication.

\subsection{CMDs and Sources with Excess Infrared Emission}
\label{CMDs}

\begin{figure*}[h]
    \centering
    \includegraphics[width=0.45\textwidth]{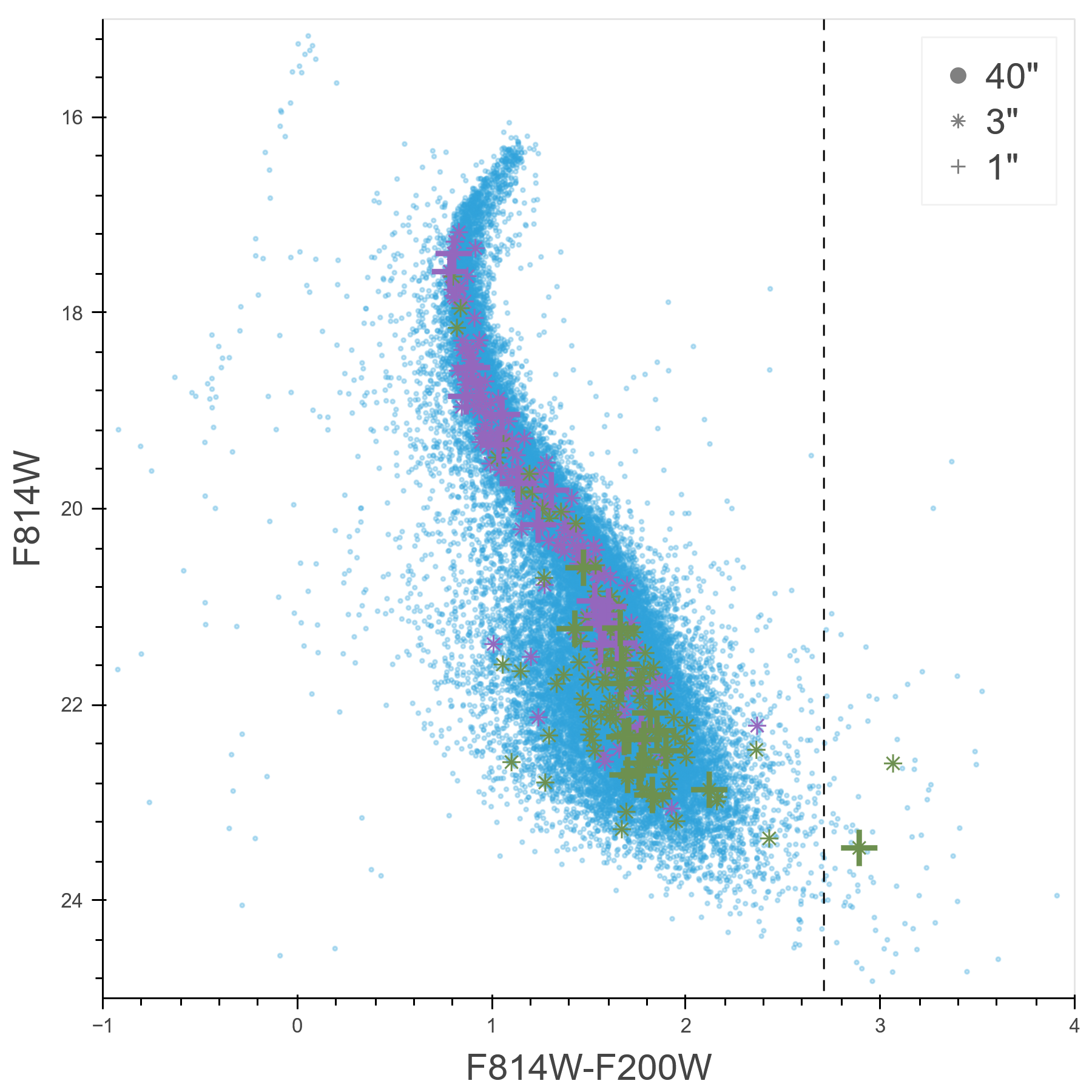}
    \includegraphics[width=0.45\textwidth]{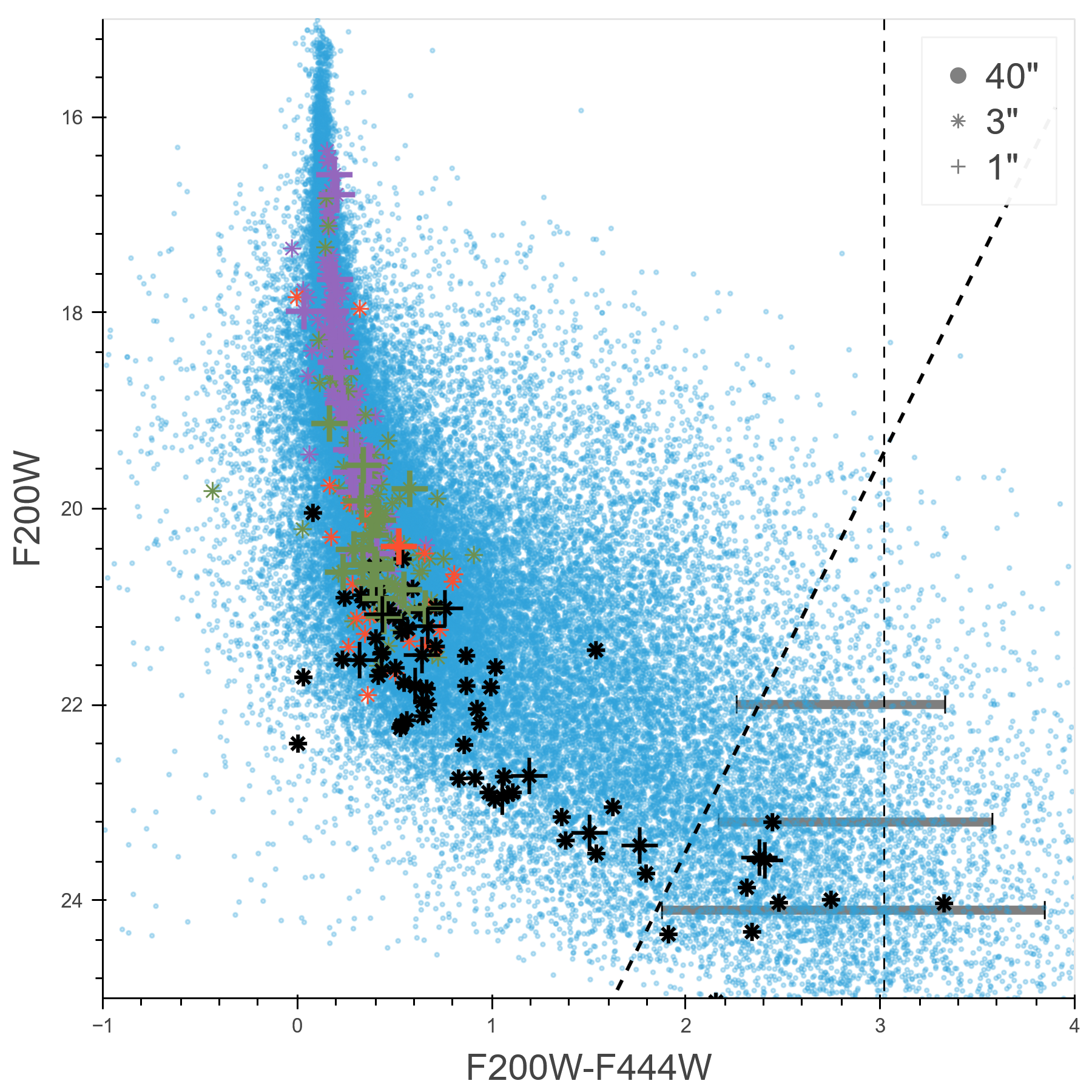}
    \includegraphics[width=0.45\textwidth]{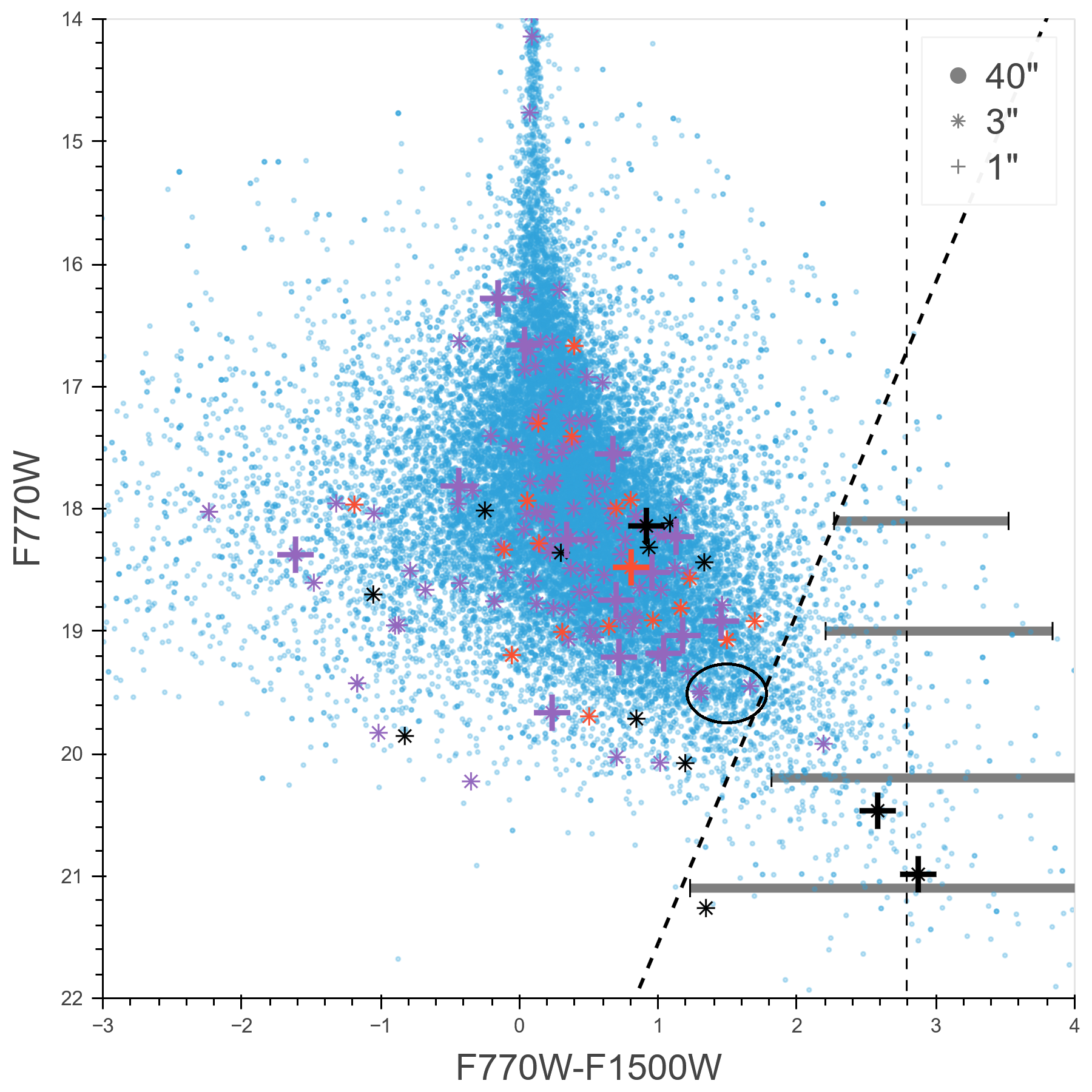}
    \caption{HST-JWST CMDs of sources within $r<1\arcsec$, $r<3\arcsec$, and $r<40\arcsec$ of the cluster center. Blue sources are all sources appearing in the $r<40\arcsec$ region. Marker colors indicate which catalogs the sources appear in the $r<3\arcsec$ regions: green sources appear in oMEGACat, NIRCam, and MIRI catalogs, purple sources in oMEGACat and NIRCam, and red sources in NIRCam and MIRI. Black sources are only detected in the catalog corresponding to the CMD where they are shown. 
    The vertical dashed lines show the bluest color expected for an IMBH accretion model of mass \qty{8.2e3}{M_\odot}, $p=0.5$, and $n=\qty{0.2}{\cm^{-3}}$ (first row of Table \ref{tab:models}). The horizontal error bars show the $2\sigma$ color error expected for sources of color 3 at different magnitudes. The diagonal dashed lines in the second and third panels are used to select ``red" sources. The SEDs of the 3 sources inside the $r<3\arcsec$ region enclosed by the ellipse in the third panel are shown in Figure \ref{fig:ngc5139_center_miri_seds}. Note that two of these sources appear close together on the CMD.}
    \label{fig:ngc5139_center_cmds}
\end{figure*}

Figure \ref{fig:ngc5139_center_cmds} shows CMDs of the cross-matched sources within $r=40\arcsec$ (small dots), $r=3\arcsec$ (asterisks), and $r=1\arcsec$ (crosses) of the cluster center. Based on SEDs calculated from P+21 for a weakly accreting IMBH (see Section \ref{Discussion} and Fig. \ref{fig:IMBH_SEDs}), we expect an IMBH to appear very red compared to the main sequence in all CMDs. The vertical dashed lines thus show the bluest color expected for an IMBH of mass \qty{8.2e3}{M_\odot}, $p=0.5$, and $n=\qty{0.02}{\cm^{-3}}$ (first row of Table \ref{tab:models}). Using artificial star injections in DOLPHOT (see Section \ref{upper limits} for details), artificial stars with color value of 3 in F200W-F444W and F770W-F1500W at different magnitudes were injected inside the $2''\times2''$ region at the cluster center. We computed 2 and 98 percentile errors in these colors from distributions of differences between the injected star color and recovered star color. Several of these error bars are shown in the corresponding CMDs. As artificial star injections involving multiple instruments are more complicated, we do not produce these errors for the F814W vs.\ F814W-F200W CMD, instead, we only discuss the 4 obvious red outliers.

While most sources within the $r<3\arcsec$ region lie on the main sequence, and well blue-ward of the vertical line, some faint sources lie close to the vertical line in each CMD. However, by carefully inspecting the images, we find that all of these sources are spurious, suffer from contamination, or are blended MIRI sources with contributions from multiple stars resolved in the HST and NIRCam images. In fact, the CMDs show no sources that consistently become brighter with increasing wavelength, as would be expected for an isolated IMBH based on the SED models.

\begin{figure*}[h]
    \centering
    \includegraphics[width=0.45\textwidth]{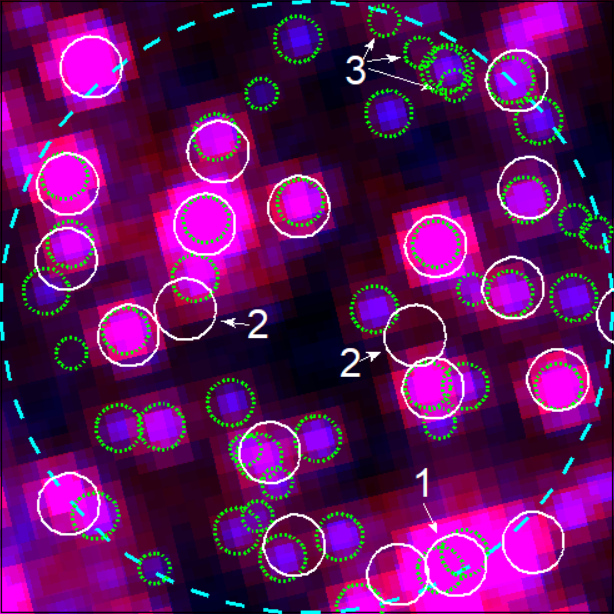}
    \includegraphics[width=0.45\textwidth]{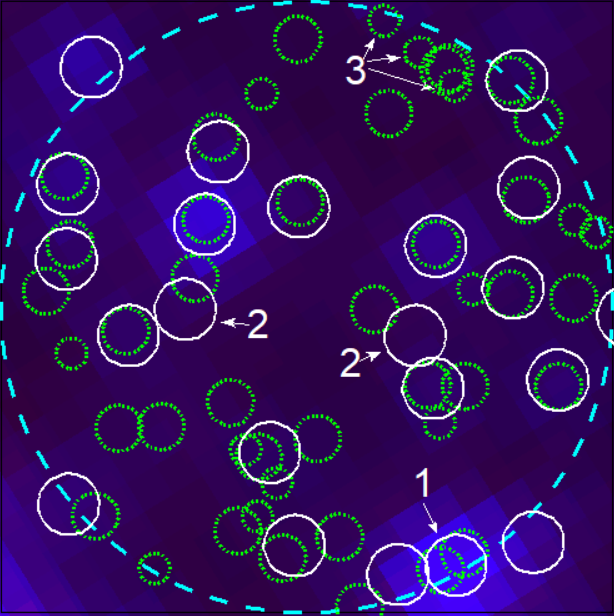}
    \includegraphics[width=0.45\textwidth]{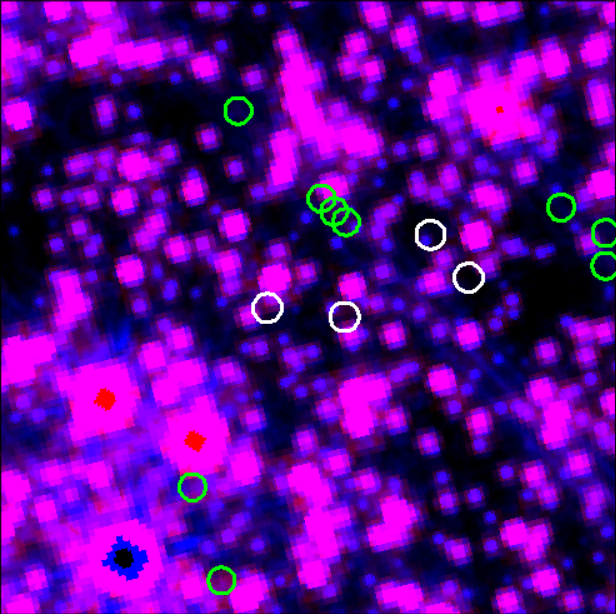}
    \includegraphics[width=0.45\textwidth]{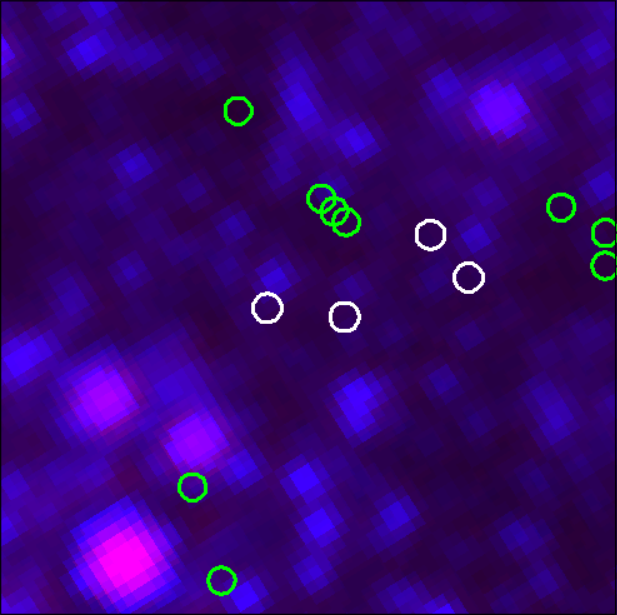}
    \caption{Detected NIRCam and MIRI sources within the central $r<1\arcsec$ region are shown by circles in the top two panels. The left (right) panel uses the NIRCam F200W, F444W (MIRI F770W, F1500W) images for the blue and red channels respectively. Green (white) circles show NIRCam (MIRI) sources. The smaller green circles show fainter NIRCam sources without an oMEGACat counterpart. Numbered labels are placed next to sources discussed in the text. The two bottom panels show the same images as top row but for the larger $r<3''$ region, with circles showing sources with apparent red excess in the NIRCam and MIRI CMDs}, as defined in Section \ref{CMDs}.
    \label{fig:ngc5139_center_jwst_sources}
\end{figure*}

To complement the CMDs, we show images of all detected NIRCam (green circle) and MIRI (white circle) sources within the central $r<1\arcsec$ region in the top two panels of Figure \ref{fig:ngc5139_center_jwst_sources}. The NIRCam sources matched to MIRI sources are generally brighter stars, but even for these relatively brighter sources, the separation between the matches can be larger than $0.1\arcsec$. We attribute this to the larger MIRI PSF, and contamination from nearby stars. Sometimes, multiple sources resolved in NIRCam are blended into a single MIRI source, as can be seen below the label ``1", where two NIRCam sources were matched to a single blended MIRI source that is significantly offset from both of them. 

In the bottom two panels of Figure \ref{fig:ngc5139_center_jwst_sources}, ``red" sources that lie within the range of the horizontal error bars in the CMDs are shown. Green (white) circles mark sources appearing in the F200W vs. F200W-F444W (F770W vs. F770W-F1500W) CMD. As no source within 3\arcsec of the center appears sufficiently red in multiple CMDs, it is unlikely any of these sources can be an isolated IMBH according to the P+21 radiative model described in Section \ref{Discussion}. Nevertheless, it is possible that the IMBH is only detected in a subset of these filters. We thus discuss these ``red" sources appearing in each CMD in detail below.

\begin{figure*}[h]
    \centering
    \includegraphics[width=0.45\textwidth]{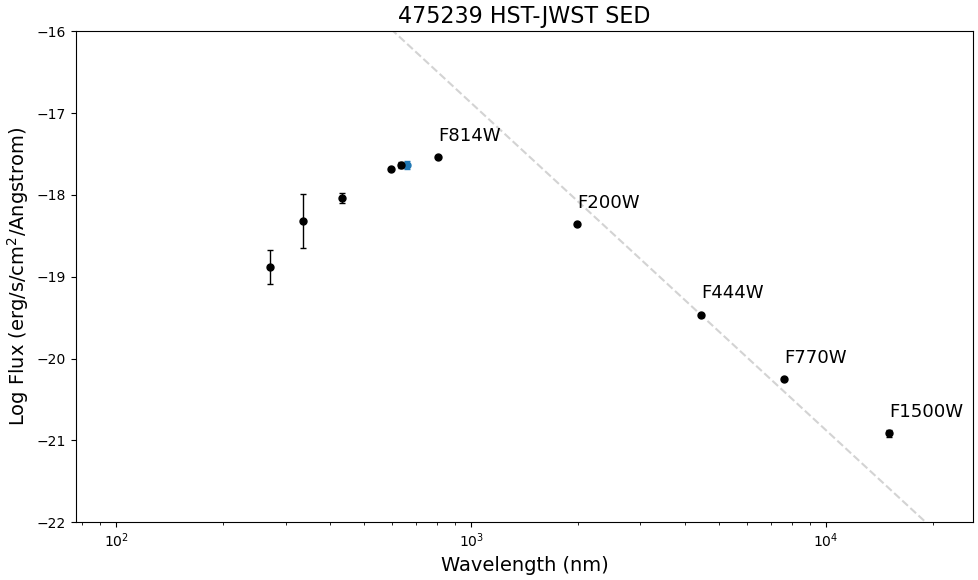}
    \includegraphics[width=0.45\textwidth]{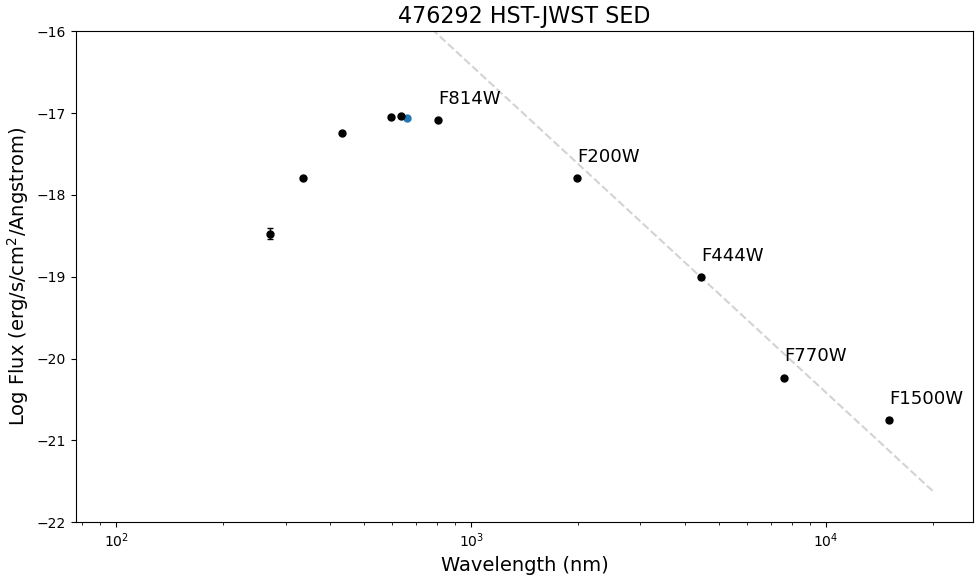}
    \includegraphics[width=0.45\textwidth]{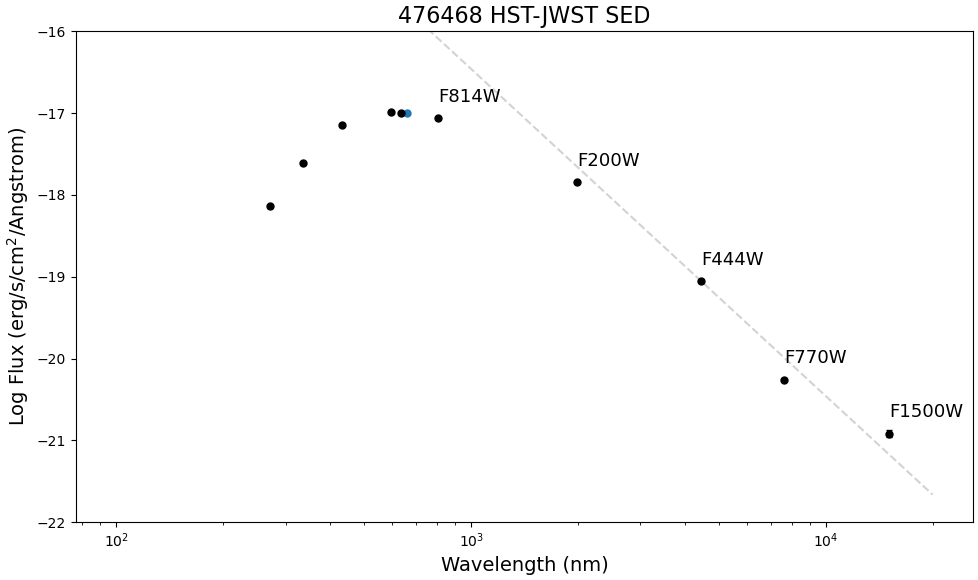}
    \includegraphics[width=0.45\textwidth]{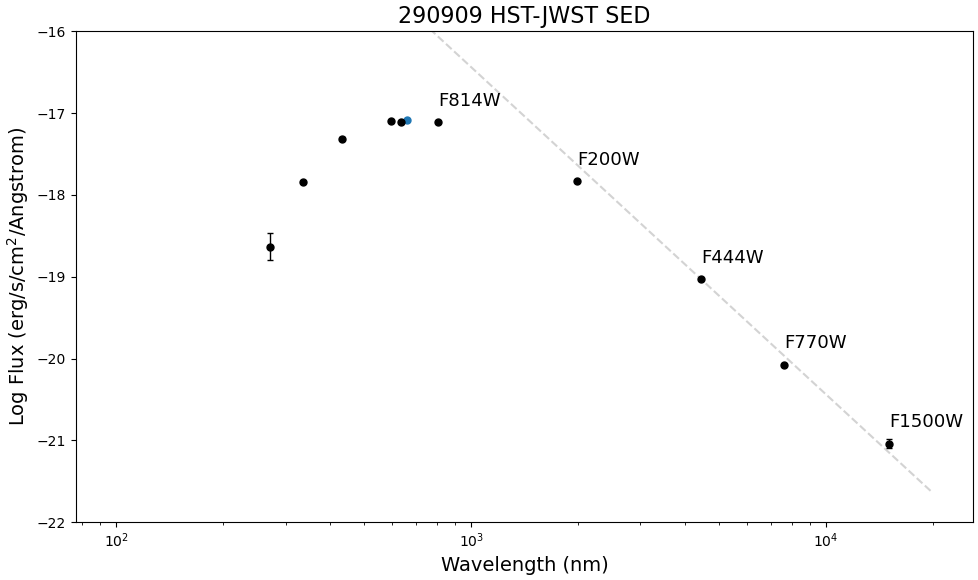}
    \caption{SEDs of three oMEGACat-NIRCam sources with MIRI counterparts showing apparent F1500W excess in the third panel of Figure \ref{fig:ngc5139_center_cmds}, enclosed by an ellipse. The last panel shows the SED of a more isolated star of similar brightness, that does not have the F1500W excess. The dashed gray lines shows power laws with index -4, normalized to the F444W band, expected for a thermal source with a Rayleigh-Jeans tail. The purely statistical uncertainties in the JWST fluxes produced by DOLPHOT appear small, but do not account for crowding. }
    \label{fig:ngc5139_center_miri_seds}
\end{figure*}

All four ``red" sources in the F770W vs.\ F770W-F1500W CMD are likely spurious sources, caused by contamination from nearby MIRI sources. The reddest two are shown in the top two panels of Figure \ref{fig:ngc5139_center_jwst_sources} next to the ``2" labels. The three sources enclosed by the ellipse in the CMD are the reddest sources that appear to be real counterparts to NIRCam sources in the images. Despite a small apparent excess in F1500W, their SEDs, shown in Figure \ref{fig:ngc5139_center_miri_seds}, are otherwise stellar-like, and thus cannot represent an isolated IMBH. If a star is close to, and therefore unresolved from an IMBH, and dominates the combined spectrum blue-ward of F1500W, then it is possible that IMBH contribution is seen as excess in F1500W only. However, the apparent excess in F1500W is more likely due to contamination by nearby stars.

In the F200W vs.\ F200W-F444W, most sources with color within the range of the horizontal error bars lie on diffraction spikes from a bright star, three of which are visible above the label ``3" in Figure \ref{fig:ngc5139_center_jwst_sources}. These sources are thus most likely spurious. The remaining ``red" sources appear to also be significantly affected by contamination.

The four sources with red excess in the F814W vs. F814W-F200W CMD are real sources that suffer from contamination. One is contaminated by the same diffraction spike as the three spurious sources in the F200W vs.\ F200W-F444W CMD.

In summary, no reliably detected source within the $<3\arcsec$ region shows red excess in multiple colors. Even if some sources appear to be red in one color, there is not enough evidence to argue that any source is an isolated IMBH. 

\section{Upper limits using artificial source injections}
\label{upper limits}

To infer the limiting magnitude, below which sources are still able to be reliably detected/recovered in the cluster center, we used artificial star injections with DOLPHOT. We created two separate artificial star catalogs, one for each detector containing the two relevant filters (i.e., F200W and F444W for NIRCam, F770W and F1500W for MIRI). Since we do not know the exact position of the IMBH, the stars were injected with random positions into a 2$''$ by 2$''$ box centered on the updated position of the IMBH from H+24. We injected 56,000 stars into each filter in NIRCam and 42,000 stars into each filter in MIRI. The magnitudes ranged between 16 and 28 for NIRCam (in 200W), and between 13 and 22 for MIRI (in F770W) in steps of 0.3 magnitudes. The injected stars were given F200W-F444W and F770W-F1500W colors of 3 (i.e., approximately the same colors expected for the IMBH, see Table \ref{tab:models}). After applying the same cuts as we used for our real photometric catalogs, 34,191 stars remained in NIRCam and all stars remained in MIRI (as no cuts were applied to MIRI) and were considered for the completeness calculation. It is important to note that the stars are injected one-by-one to avoid significantly overcrowding the photometry (see e.g. \citealt{weisz_jwst_2024}). 

\begin{figure*}[h]
\begin{center}
\includegraphics[width=1.0\textwidth]{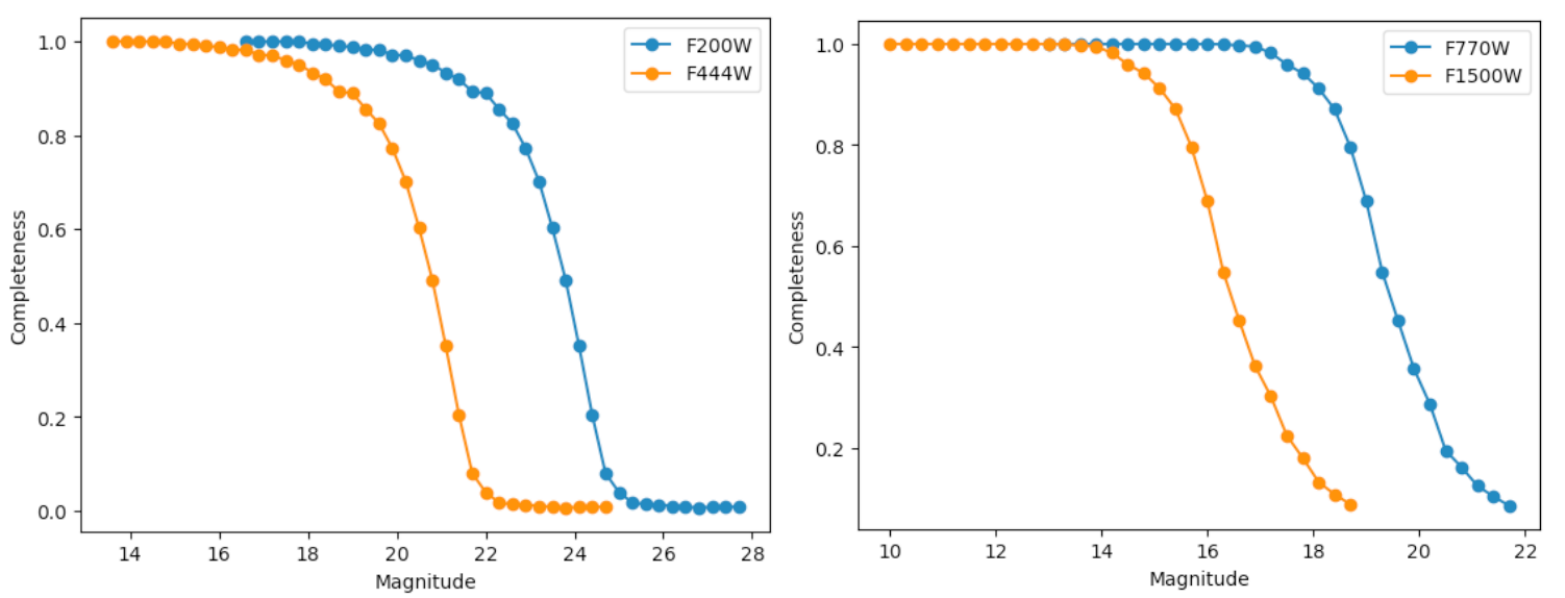}
\caption{Completeness versus magnitude for artificial star injections into the NIRCam (F200W/F444W) and MIRI (F770W/F1500W) images.
}
\label{fig:art_stars}
\vspace{-0.0cm}
\end{center}
\end{figure*}

The completeness versus magnitude plots from these artificial star tests are shown in Figure \ref{fig:art_stars}. The amount of crowding, and thus the limiting detectable magnitude, is heavily dependent on the exact spatial position of the injected artificial star (or IMBH when considering the upper-limits) even within the 2$''$ by 2$''$ box region, as only relatively bright artificial stars can be recovered close to bright, real stars\footnote{We also ran the artificial star tests in a relatively star free ``good'' region and in a ``bad'' region with one bright and several fainter stars surrounding it, to get a sense how large of an impact the spatial location of the injected star has on its chance to be recovered. We found that the 50$\%$ completeness value was about two magnitudes deeper for the ``good`` region compared to the ``bad'' region, and about one magnitude deeper compared to the results shown in Figure \ref{fig:art_stars}.}. Therefore, we use the completeness fractions of 99.7\%, 95\%, and 68\% to show the corresponding limiting fluxes in Figure \ref{fig:IMBH_SEDs} by  horizontal black bars. These fluxes, magnitudes, and corresponding observed luminosities are listed in Table \ref{tab:limits}. We note that the first two points in both the F200W and F444W completeness curves were removed, as the stars were too bright, at 13 and 13.3 mag, leading to saturation and poor photometric quality.

We also use the artificial sources to estimate the error in the color for the recovered sources as a function of the recovered source magnitude, which we then convert to flux. This is done by calculating the 2 and 98 percentiles of the recovered color distribution at each input magnitude and adding these to the mean recovered color offset (from the input color of 3) in quadrature. The fluxes corresponding to an uncertainty of 1 in the recovered F200W-F444W and F770W-F1500W colors are shown in Figure \ref{fig:IMBH_SEDs} by red horizontal bars.

\begin{deluxetable*}{lcccc}
\tablecaption{Photometric Upper Limits. \label{tab:limits}}
\tablewidth{0pt}
\tablehead{
\colhead{Filter} & \colhead{Completeness/Color Unc.} & \colhead{Magnitude Limit} & \colhead{Flux Density Limit} & \colhead{Luminosity (at 5.49 kpc)}\\
\colhead{} & \colhead{} & \colhead{(Vega mag)} & \colhead{($10^{-19}$ erg s$^{-1}$ cm$^{-2}$ \AA$^{-1}$)} & \colhead{($10^{30}$ erg s$^{-1}$)}
}
\startdata
F200W & 99.7\% & $>17.8$ & $<44$ & $<75$ \\
      & 95\% & $>20.8$ & $<2.8$ & $<4.8$ \\
      & 68\% & $>23.2$ & $<0.30$ & $<0.51$ \\
      & $\delta_{F200W-F444}=1$ & $>24.1$ & $<0.13$ & $<0.22$ \\
\hline
F444W & 99.7\% & $>14.8$ & $<37$ & $<150$ \\
      & 95\% & $>17.8$ & $<2.3$ & $<9.2$ \\
      & 68\% & $>20.2$ & $<0.25$ & $<1.0$ \\
       & $\delta_{F200W-F444}=1$ & $>21.1$ & $<0.10$ & $<0.4$ \\
\hline
F770W & 99.7\% & $>16.6$ & $<0.82$ & $<4.4$ \\
& 95\% & $>17.5$ & $<0.36$ & $<1.9$ \\
      & 68\% & $>19.0$ & $<0.09$ & $<0.49$ \\
      & $\delta_{F770W-F1500}=1$ & $>20.2$ & $<0.027$ & $<0.15$ \\
\hline
F1500W & 99.7\% & $>13.6$ & $<1.17$ & $<11$ \\
& 95\% & $>14.5$ & $<0.51$ & $<4.9$ \\
       & 68\% & $>16.0$ & $<0.12$ & $<1.2$ \\
       & $\delta_{F770W-F1500}=1$ & $>17.2$ & $<0.042$ & $<0.41$ \\
\enddata
\tablecomments{Detection upper limits for the four JWST filters. See Section \ref{upper limits} for details. Luminosity calculated as $L=4\pi d^2 F_\lambda \Delta\lambda$, where $\Delta\lambda$ is the width of the filter.}
\end{deluxetable*}

\section{Discussion}
\label{Discussion}

Based on motions of fast stars in the core of $\omega$ Cen,  H+24 found that the most plausible range of IMBH masses is 39,000$M_{\odot}$-47,000$M_{\odot}$, with a lower limit of 8,200$M_{\odot}$. This limit increases to 21,100$M_{\odot}$ (at 99\% confidence) when the (unknown) line-of-sight distances and acceleration limits of the fast stars are included in addition to their velocities. In the earlier work of \cite{van_der_marel_new_2010} the dynamical limit on the IMBH mass was found to be $1.8\times10^{4}M_{\odot}$ (at $3\sigma$ confidence).

We compare these dynamical mass estimates with those allowed by our JWST upper limits, as well as the previously published X-ray and radio limits on a point source in the center of $\omega$ Cen \citep{haggard_deep_2013,tremou_maveric_2018,mahida_no_2025}. In order to compute the SED of the putative IMBH, we use the radiative model (and the corresponding Python code) from  P+21. The radiative model has two input parameters: the accretion rate $\dot{m}_0$ in Eddington units at the Schwarzschild radius $R_S=2GM_{\rm BH}/c^2$, and the BH mass $M_{\rm BH}$. The accretion rate as a function of radius, in Eddington units, is then $\dot{m}=\dot{M}/\dot{M}_{\rm Edd}=\dot{m}_0(R/R_S)^p$ with $0\le p \le 1$. The term $(R/R_S)^p$ controls how much mass outflows as matter is accreted down to $R_S$. With these inputs, the code calculates the broadband spectrum from radio to X-rays. In order to connect $\dot{m}_0$ to the properties of the intracluster medium, we express it through the Bondi accretion rate $\dot{M} =\dot{M}_{\rm Bondi}$ at $R=R_{\rm Bondi}$:

\begin{align}
\label{eq:mdot}
 \dot{m}_0 &= \zeta^{p} \frac{\dot{M}_{\rm Bondi}}{\dot{M}_{\rm Edd}} \\
 &= \zeta^{p} \frac{ \eta GM_{\rm BH} nc \sigma_{\rm T}}{(v_{\rm BH}^2+c_s^2)^{3/2}}, \nonumber
\end{align}

Where the Eddington accretion rate is

\begin{align}
\dot{M}_{\rm Edd} &= \frac{L_{\rm Edd}}{ \eta c^2} \\
&= \frac{4 \pi G M_{\rm BH} m_p}{ \eta \sigma_{\rm T}c} \nonumber \\ 
&= 2.1\times10^{-4} \left(\frac{M_{\rm BH}}{10^4M_{\odot}}\right) \left(\frac{\eta}{0.1}\right) ~M_{\odot}~{\rm yr}^{-1}. \nonumber 
\end{align}

where $\eta = 0.1$ is the maximum radiative efficiency as in Equation A3C of P+21. Note that in a radiatively inefficient regime, relevant for an IMBH in a low density environment, the actual radiative efficiency is much lower (see below). 

The Bondi accretion rate \citep{bondi_spherically_1952, shima_hydrodynamic_1985} is

\begin{align}
\dot{M}_{\rm Bondi} &= \frac{4\pi G^2 M_{\rm BH}^2 n m_p}{(v_{\rm BH}^2+c_s^2)^{3/2}} \\
&= 2\times10^{18} \left(\frac{M_{\rm BH}}{\qty{1e4}{M_{\odot}}}\right)^2\left(\frac{n}{\qty{0.1}{\per\cm^3}}\right)\left(\frac{T}{\qty{1e4}{\kelvin}}\right)^{-3/2}~{\rm g~s}^{-1} \nonumber \\ 
&{\rm for}~v_{\rm BH}\ll c_s, \nonumber
\end{align}

And $\zeta^p$ is a factor that lowers the Bondi accretion rate, with $\zeta$ defined as the ratio

\begin{align}
\zeta &= \frac{R_{\rm S}}{R_a} \\
&= \frac{v_{\rm BH}^2+c_s^2}{c^2}\approx 1.6 \times10^{-9} \left(\frac{T}{\qty{1e4}{\kelvin}}\right) \nonumber \\
&{\rm for}~v_{\rm BH}\ll c_s, \nonumber
\end{align}

with Bondi gravitational capture radius, or Bondi radius for short.\footnote{The gravitational capture radius of 0.59 pc corresponds to the angular size of $\approx23\arcsec$ at $d=5.49$ kpc.}

\begin{align}
R_a &= \frac{2GM_{\rm BH}}{(v_{\rm BH}^2+c_s^2)} \\ 
&\approx0.59\left(\frac{M_{\rm BH}}{\qty{1e4}{M_{\odot}}}\right)\left(\frac{T}{\qty{1e4}{\kelvin}}\right)^{-1}~{\rm pc} \nonumber \\
&{\rm for}~v_{\rm BH}\ll c_s, \nonumber
\end{align}

and isothermal sound speed for intracluster gas with mean molecular weight $\mu=0.59$ (see \citealt{fall_theory_1985})

\begin{align}
c_s &= \left(\frac{ k_{\rm B} T}{\mu m_p}\right)^{1/2} \\
&= 12\left(\frac{T}{\qty{1e4}{\kelvin}}\right)^{1/2}\left(\frac{\mu}{0.59}\right)^{-1/2}~{\rm km}~{\rm s}^{-1}, \nonumber 
\end{align}

From Equation (1), the intracluster medium number density $n$ can be expressed through $M_{\rm BH}$, $\dot{m}_0$, and $p$ as 

\begin{align}
\label{eq:density}
n &\approx 9\times10^{-3} (1.6 \times 10^{-9})^{\frac{p}{0.3}} \nonumber\\
&\left(\frac{M_{\rm BH}}{\qty{1e4}{M_{\odot}}}\right)^{-1}\left(\frac{\dot{m}_0}{3\times10^{-8}}\right)~{\rm cm}^{{-3}},
\end{align}

for $T=10,000$ K, $\eta=0.1$, and $v_{\rm BH}\ll c_s$.

Note that lower values of $p$ result in higher $\dot{m_0}$, and at $p=0$ the BH is accreting at the Bondi-Hoyle rate. When $p>0$, part of the in-falling matter is assumed to be ejected (outflow), thus, reducing the accretion rate (see \cite{yuan_hot_2014} for a more detailed discussion). Since the exact value of $p$ is unknown, we fix it for the rest of the calculations at $p=0.3$, which has some observational support as it has been suggested for Sgr A$^{\ast}$ \citep[][where it is called $s$]{yuan_nonthermal_2003, yuan_hot_2014}. This value corresponds to $\sim 0.2\%$ of the accreted matter at the Bondi radius reaching the Schwarzschild radius. 

To compute the model SEDs, we need to estimate $\dot{m}_0$ for the expected range of $M_{\rm BH}$ and plausible properties of the intracluster gas within $\omega$ Cen. If a BH is formed and retained in the cluster's core, its velocity, $v_{\rm BH}$, is expected to be smaller than the GC escape velocity ($\sim60$ km s$^{-1}$ for $\omega$ Cen; \citealt{gnedin_unique_2002,haberle_fast-moving_2024}). For the large IMBH masses considered in this paper, $v_{\rm BH}$ should be much lower, and hence we neglect the IMBH velocity in comparison with the sound speed $c_s$. The atomic hydrogen number density, $n$, in the core of $\omega$ Cen is not precisely known. Based on the stellar mass-loss rates discussed in \cite{mcdonald_giants_2009} and \cite{wang_omegacat_2025}, a plausible range appears to be $n=$0.01-0.2 cm$^{-3}$ with most likely values being toward the middle of this range. Based on non-detection of sodium absorption lines, \cite{wang_omegacat_2025} also set an upper limit to the intracluster atomic gas column density in $\omega$ Cen of $\lesssim\qty{2.17e18}{\cm^{-2}}$, which corresponds to a space density of $\lesssim\qty{0.045}{\cm^{-3}}$ over a uniform 15 pc column, compatible with the density range above. Additionally, \cite{wang_omegacat_2025} used dispersion measures (DMs) of pulsars in $\omega$ Cen and found a significantly higher upper limit of ionized gas density, $n_e\lesssim2$ cm$^{-3}$, which, however, could be largely due to DM variations foreground to the cluster. The temperature of the intracluster gas is expected to be in the $10,000$-$20,000$ K range \citep{haggard_deep_2013}. For simplicity, in addition to fixed $p=0.3$, we assume $T=10,000$ K everywhere below, and only change $n$ to vary $\dot{m}_0$. 

\begin{figure*}[h]
\begin{center}
\vspace{-0.8cm}
\includegraphics[width=0.7\textwidth]{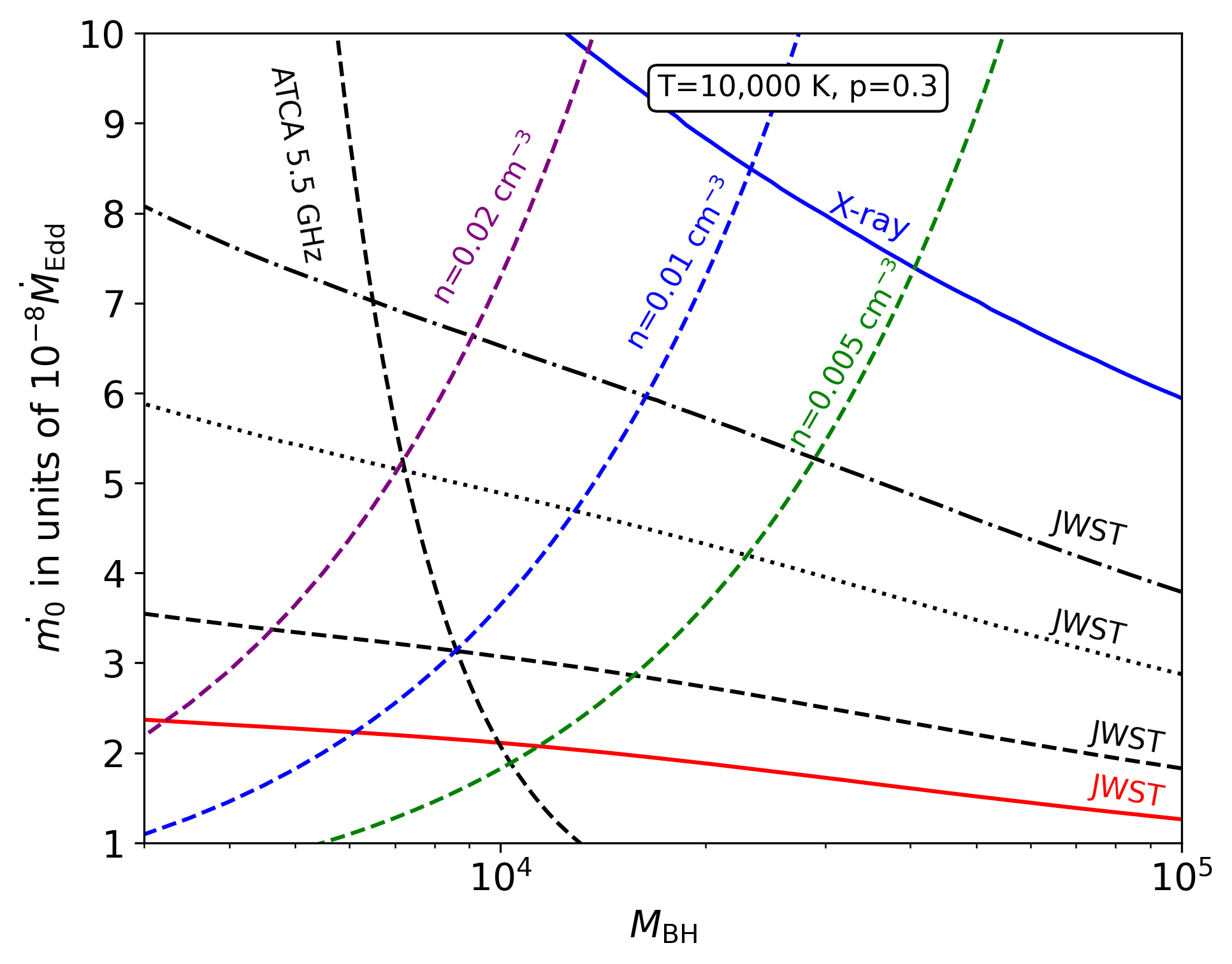}
\vspace{-0.0cm}
\caption{ Contour plot of IMBH mass $M_{\rm BH}$ vs mass accretion rate at the Schwarzschild radius $\dot{m}_0$. The contours of constant spectral flux at the central frequency of F770W (labeled ``JWST'') correspond to flux values for different completeness fractions shown in Table \ref{tab:limits}. The contours of constant flux corresponding to X-ray and radio upper limits (see text) are also shown. Constant number density contours (see equation \ref{eq:density}) are calculated for $T=10,000$ K and $p=0.3$.  
}
\label{fig:m_mdot}
\vspace{-0.0cm}
\end{center}
\end{figure*}

\begin{figure*}[h]
\begin{center}
\vspace{-0.8cm}
\includegraphics[width=1.0\textwidth]{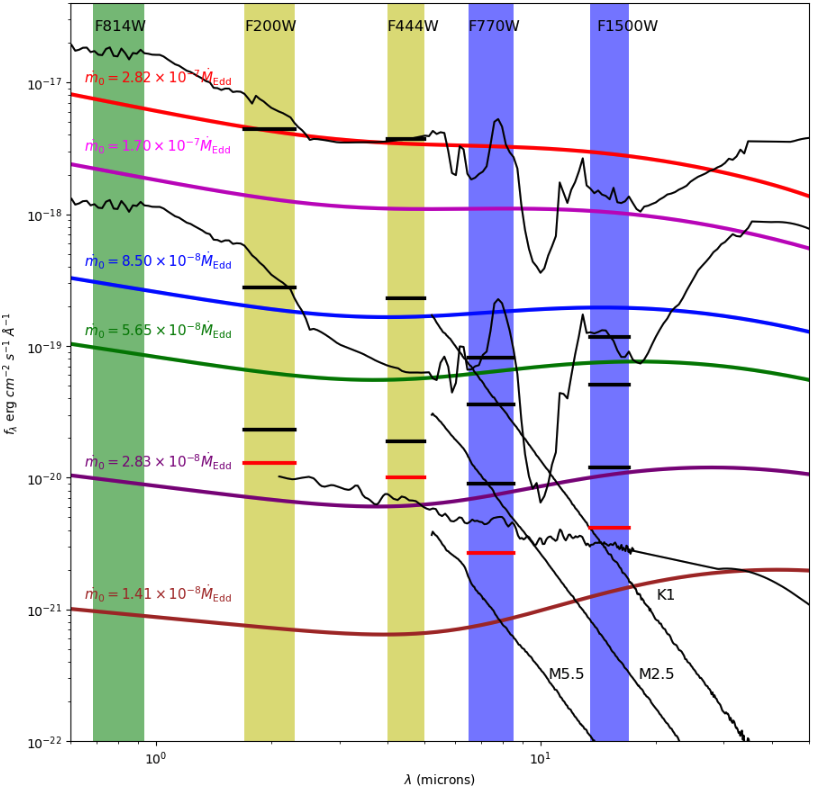}
\vspace{-0.0cm}
\caption{
The thick colored curves are IMBH SEDs, computed with the radiative model from P+21, parametrized by  accretion rate $\dot{m}_0$ for fixed $M_{BH}=8,200M_{\odot}$, $T=10,000$ K, and $p=0.3$ (see text). The bottom three curves correspond to the three constant values of intracluster gas number density shown in Figure \ref{fig:m_mdot}. The thinner black solid curves are examples of SEDs of typical cool low-mass dwarfs (spectral types M5, M2, and K1 from bottom to top) at the distance of $\omega$ Cen, and IR-luminous AGNs which can exhibit broad silicate absorption features. Vertical colored bands correspond to MIRI's F1500W and F770W, NIRCam's F444W and F200W, as well as HST's F814W (from right to left). The horizontal black bars in each band correspond to fluxes at which 99.7\%, 95\%, and 68\% (top to bottom) of injected artificial sources are recovered (see Section \ref{upper limits}). The red horizontal bars indicate the fluxes at which the color measurement uncertainty (inferred from comparing the recovered and original color of artificial sources) becomes equal to 1. See also  Tables \ref{tab:limits}    and \ref{tab:models}. 
}
\label{fig:IMBH_SEDs}
\vspace{-0.0cm}
\end{center}
\end{figure*}

Figure \ref{fig:m_mdot} shows contours of constant spectral flux (labeled ``JWST'') at the central frequency of F770W (shown in Table \ref{tab:limits}, and corresponding to the four flux values shown in Figure \ref{fig:IMBH_SEDs} by horizontal bars) in the plane of $M_{\rm BH}-\dot{m}_0$. In addition, we also plot contours corresponding to radio ($\approx 3.3$ $\mu$Jy at 5.5 GHz \citep{mahida_no_2025}) and X-ray (0.07 nJy at 1 keV; \citealt{haggard_deep_2013}) spectral flux limits computed with the same radiative model. In order to show what range of $\dot{m}_0$ corresponds to plausible conditions in $\omega$ Cen, we also plot contours (shown in different colors) corresponding to several fixed values of $n$ for $T=10,000$ K and $p=0.3$.

From Figure \ref{fig:m_mdot}, one can see that the ATCA radio flux limit is more restrictive than the JWST data for $M_{\rm BH}\gtrsim6,000M_{\odot}$. For lower masses, neither JWST nor radio observation are restrictive for $\dot{m}_0\lesssim2\times 10^{-8}$, for the chosen $T$ and $p$ values.

We note that, based on P+21, neither radio nor JWST observations rule out higher IMBH masses if the accretion rate is significantly below the $\dot{m}_0$ range shown in Figure \ref{fig:m_mdot}. However, for the chosen $T=10,000$ K and $p=0.3$, such low $\dot{m}_0$ values require an unrealistically low gas number density ($<0.001$ cm$^{-3}$). This discrepancy is alleviated if larger values of $p$ and/or higher gas temperatures are used. For the $M_{\rm BH}$ and $\dot{m}$ ranges considered here, the X-ray upper limit is much less constraining than the JWST and radio limits.

Figure \ref{fig:IMBH_SEDs} shows model IMBH SEDs calculated for $M_{\rm{BH}}=8,200M_{\odot}$ and several $\dot{m}_0$ values. The three bottom curves correspond to the three density contours shown in Figure \ref{fig:m_mdot}, where Equation \ref{eq:mdot} is used to relate $\dot{m}_0$ and $n$. Figure \ref{fig:IMBH_SEDs} also shows representative spectra of galaxies and AGN, as well as spectra of low mass dwarf stars which are abundant in $\omega$ Cen. The HST and JWST bands that we used for building the CMDs are shown by vertical shaded bands together with the completeness and detection limits discussed in Section \ref{upper limits}. 

We briefly discuss previous limits on the IMBH mass and radiative efficiency imposed by other constraints on its emission. \cite{haggard_deep_2013} used a limit on the observed X-ray flux to constrain the IMBH mass to be below $M_{BH}\lesssim (3.7-5.3)\times 10^3M_{\odot}$, or, even $M_{BH}\lesssim (120-200)\times 10^3M_{\odot}$, depending on different assumptions about radiative efficiencies, which differ from our direct approach to calculating the IMBH SED. \cite{tremou_maveric_2018} reported an upper limit of $M_{BH}<\qty{1000}{M_\odot}$ based on radio observations with ATCA, which is nearly an order of magnitude lower than what follows from the radiative model we use. Their calculations rely on the empirical relationship known as the fundamental plane of BH activity, which relates $M_{BH}$, X-ray, and radio luminosity, in the form presented in \cite{miller-jones_absence_2012,plotkin_using_2012}. The relationship has a large scatter \citep{gultekin_fundamental_2019} which also translates into a large uncertainty on the radiative efficiency \citep[][which refers to it as "accretion efficiency"]{mahida_no_2025}. Compared to these works, our analysis uses the radiative model presented in P+21 to calculate the IMBH spectrum and infer limits on its mass, while still relying on some assumptions about the intracluster medium, which is used to calculate the mass accretion rate. Additionally, the P+21 model explicitly accounts for mass outflow $\zeta^p$ (in Equation \ref{eq:mdot}), with our chosen value of $p=0.3$ corresponding to $\sim 0.2\%$ of the accreted matter at the Bondi radius reaching the Schwarzschild radius. Other works, including \cite{mahida_no_2025} and one approach discussed in \cite{haggard_deep_2013}, assumed a simple mass accretion rate of 3\% of the Bondi rate without outflows.

Using the P+21 radiative model, the radio and JWST flux limits exclude the most likely mass $M_{BH} \sim {40,000}{M_\odot}$ suggested by H+24, and the $\sim {50,000}{M_\odot}$ produced by the simulations in \cite{prieto_growing_2025}, for the range of accretion rates based on plausible assumptions we made. The firm {\bf H+24's} $M_{BH}$ lower limit of $\sim {8,200}{M_\odot}$ is excluded for accretion rates $\dot{m}_0\gtrsim3\times10^{-8}$ by the JWST IR data (at 68\% flux completeness limit) and by the \cite{mahida_no_2025} ATCA radio data (at $3\sigma$ flux limit). With $M_{BH} = {8,200}{M_\odot}$ and $\dot{m}_0= 3\times10^{-8}$, and using the SEDs from the P+21 model (see examples in Figure \ref{fig:IMBH_SEDs}), the radiative efficiency ($\epsilon = \frac{L}{\dot{M}_{\rm Bondi}c^2}$) in the F770W band is $3\times 10^{-9}$ while the bolometric radiative efficiency (from 100 Hz to $10^{22}$ Hz) is $6\times 10^{-7}$.  This work provides the first estimate of radiative efficiency in IR.
The bolometric radiative efficiency can be scaled to the Eddington luminosity instead $\epsilon_{\rm bol} = L_{\rm bol}/L_{\rm Edd}$, obtaining a value of $8\times 10^{-11}$. This value is the same order of magnitude as those reported in \cite{haggard_deep_2013} ($\epsilon_{\rm bol}<1.4\times 10^{-11}$) and \cite{tremou_maveric_2018} ($\epsilon_{\rm bol}\lesssim2\times 10^{-11}$), which were estimated using different assumptions. 

The discrepancy between our mass limits and the most likely $M_{BH}$ inferred by H+24 can be resolved by a higher outflow rate of matter between the Bondi radius and innermost stable circular orbit governed by $p$ in Equation \ref{eq:mdot}. If $p\gtrsim0.4$, the current detection limits from JWST are not constraining for the $M_{\rm{BH}}$ and $\dot{m}_0$ range considered here. It is also possible that $n$ is significantly lower, or $T$ is significantly higher, in the vicinity of the IMBH than we assumed, since these parameters are not currently well constrained. Additionally, it is possible that a cluster star is either coincidentally projected close to the IMBH, or orbiting it\footnote{If the orbit is tight enough, it may not be possible to detect such a star with PM measurements despite the very fast motion. In this case, radial velocity measurements are necessary, which can be challenging for faint stars.}. In this case, the bluer bands can be dominated by the stellar emission, while the IMBH emission may only show up in the reddest bands. While a few sources show an excess in F1500W, we do not find them to be convincing enough for further discussion here (see Section \ref{CMDs}). Additionally, an IMBH located near a bright star would need to have a higher infrared flux, corresponding to the higher completeness fraction lines in Figure \ref{fig:m_mdot}, to be detected despite the contamination from the star. Finally, \cite{su_chandra_2022} argues that stellar wind from a bound star orbiting the IMBH may dominate the accretion material, and since the wind has much higher velocities than the ICM, the accretion rate would be much lower than the Bondi rate. For a separation of 100 AU, a MS star orbiting a $\qty{10000}{M_\odot}$ IMBH has a period of $\sim 10$ yr. If IMBH emissions are not detectable due to such a companion, PM measurements in the near future will nevertheless be able to detect its acceleration and confirm the presence of the IMBH. 

\section{Conclusions}
\label{conclusions}

We analyzed JWST NIRCam and MIRI images of the core of $\omega$ Cen focusing on the $r<3''$ region centered at the likely IMBH location based on a study of fast moving stars in archival HST images reported in H+24. We used DOLPHOT to perform PSF-fitting photometry of stars within this region. We then compared our photometric measurements to the theoretical IMBH SEDs computed with the P+21 emission model for radiatively inefficient low-rate accretion onto a BH. Down to the completeness and detection limits set by our ability to resolve the sources and measure their colors, and barring sources that were filtered out due to artifacts in this crowded field, we do not find any sources with spectra similar to the computed IMBH spectra. In the $r<3\arcsec$ central region, we inspected all sources that appeared redder than the main sequence, but did not find any plausible IMBH candidate. 

However, our results do not exclude the existence of IMBH within the mass range suggested by H+24. For instance, with the P+21 radiative model, a choice of $p=0.3$, and $T=10,000$ K, an IMBH with $M_{\rm BH}\lesssim 10,000 M_{\odot}$ and $\dot{m}_0\lesssim 2\times 10^{-8}$ (corresponding to a plausible intracluster density) is allowed by both JWST limits and by the 5.5 GHz radio limit from \cite{mahida_no_2025}. It is worth noting that mass limits that follow from Figure \ref{fig:m_mdot} are less restrictive than the limits reported in previous studies \citep{tremou_maveric_2018, haggard_deep_2013} where different assumptions about the radiative efficiency and accretion rates were made. For $M_{\rm BH}\lesssim 6,000 M_{\odot}$, the JWST limits are more restrictive (in terms of $\dot{m}$) than the radio and X-ray limits (see Figure \ref{fig:m_mdot}). For larger BH masses, the radio limit from \cite{mahida_no_2025} is more restrictive than the JWST limits.

It is also possible that the IMBH is coincident with (e.g., projected near) a star in $\omega$ Cen, in which case the IMBH emission may only be discernible in the longest wavelength MIRI band or not at all. However, at least for larger IMBH masses, radio observations would be more restrictive in this case, assuming that nearby star is not active in radio.

In conclusion, despite the unprecedented depth and resolution that JWST offers, searching for IMBH signals in very crowded environments remains challenging. Flux limits in $\omega$ Cen strongly depend on proximity to stars, while limits on the IMBH mass also depend on the accretion model and assumptions about the mass accretion rate. Nevertheless, future JWST observations can further improve PM measurements of stars obtained from 20 years of HST observations. The depth of JWST images also allows for the discovery of new fast, but faint, stars. If these future PM measurements further substantiate the existence of an IMBH in $\omega$ Cen, limits on its emission can be used to refine models of BH emission in low accretion rate regimes.

\begin{deluxetable*}{lccccc}
\tablecaption{IMBH SED Colors for Different Filter Combinations.  \label{tab:colors}}
\tablewidth{0pt}
\tablehead{
\colhead{$\dot{m}_0$} & \colhead{$n$} & \colhead{F814W$-$F200W} & \colhead{F200W$-$F444W} & \colhead{F444W$-$F770W} & \colhead{F770W$-$F1500W} \\
 \colhead{($10^{-7})$} &  \colhead{($\rm{cm}^{-3}$)} & \colhead{(Vega mag)} & \colhead{(Vega mag)} & \colhead{(Vega mag)} & \colhead{(Vega mag)}
}
\startdata
 2.83  & 0.1 &  2.71 &  3.02 & 2.27 & 2.79 \\ 
 1.70  & 0.06 & 2.73 & 3.07 & 2.33 & 2.86 \\ 
 0.85  & 0.03 & 2.78 & 3.11 & 2.45 & 3.01 \\ 
 0.56  & 0.02 & 2.81 & 3.14 & 2.48 & 3.10 \\  
 0.28  & 0.01 & 2.88 & 3.14 & 2.54 & 3.30  \\
0.14  & 0.005 & 2.95 & 3.14 & 2.56 & 3.49 
\enddata
\tablecomments{ Calculated NIRCam and MIRI filter colors for five different IMBH models. The models are computed for $M_{\rm BH}=8,200M_{\odot}$, $p=0.3$, $T=10,000$ K, and varying accretion rate, $\dot{m}$ (scaled to the Eddington rate), due to changing intracluster gas density, $n$.}
\label{tab:models}
\end{deluxetable*}

\newpage

\appendix

\label{appendix_alignment}

Below we describe the main steps taken to match sources detected in JWST images to their oMEGACat counterparts and to align JWST images with HST images from oMEGACat. 

\begin{itemize}
    \item Propagation of oMEGACat catalog source positions to the JWST observation epoch using cluster and individual star PMs.
    \item Alignment of NIRCam source positions (from the DOLPHOT catalog) to the propagated oMEGACat source positions, using \textsc{TWEAKREG} to compute the optimal shift transformation.
    \item Alignment of MIRI source positions (from the DOLPHOT catalog) to the already corrected NIRCam source positions.
    \item Propagation of oMEGACat images to the JWST observation epoch using cluster PM.
    \item Alignment of NIRCam and MIRI images to propagated oMEGACat images, by applying the corresponding TWEAKREG shifts to adjust the WCS transformations in the original images.
\end{itemize}

A more detailed description can be found in the supplemental materials available online\footnote{\url{https://zenodo.org/records/18604036}}. This includes the Jupyter notebook which implements the described above workflow, the aligned images, DOLPHOT parameter files, and catalogs of NIRCam and MIRI sources matched to oMEGACat. 

\section{Acknowledgments}

This research is supported by Space Telescope Science Institute award JWST-GO-04343.001. J.H. acknowledges support from NASA under award number 80GSFC24M0006.

Some of the data presented in this article were obtained from the Mikulski Archive for Space Telescopes (MAST) at the Space Telescope Science Institute. The specific observations analyzed can be accessed via \dataset[doi: 10.17909/mww2-nd11]{https://doi.org/10.17909/mww2-nd11}.

We thank Dom Pesce for useful discussions on the use of the P+21 radiative model code, and the referee for constructive comments that improves the quality of the paper. 

\textit{Facility}: HST, JWST

\textit{Software}: Astropy \citep{collaboration_astropy_2013, collaboration_astropy_2022, astropy_collaboration_astropy_2018}, Matplotlib \citep{hunter_matplotlib_2007}, Numpy \citep{harris_array_2020}, IPython \citep{perez_ipython_2007}, Synphot \citep{stsci_development_team_synphot_2018}, Hvplot \citep{rudiger_holovizhvplot_2025}

\bibliographystyle{aasjournal}
\bibliography{references.bib}

\end{document}